\documentclass[RNAAS]{aastex62}

\received{January 1, 2018}
\accepted{\today}
\submitjournal{A\&A Research Note}

\shorttitle{{\it Swift} observations of asassn-18fs}
\shortauthors{Parikh \& Wijnands}
\begin{document}

\title{{\it Swift} observations of the dwarf nova asassn-18fs}

\correspondingauthor{Aastha S. Parikh}
\email{a.s.parikh@uva.nl}

\author{Aastha S. Parikh}
\affil{Anton Pannekoek Institute for Astronomy\\
 University of Amsterdam, Postbus 94249\\
  1090 GE Amsterdam, The Netherlands}

\author{Rudy Wijnands}
\affil{Anton Pannekoek Institute for Astronomy\\
 University of Amsterdam, Postbus 94249\\
  1090 GE Amsterdam, The Netherlands}

%\author{Juan V. Hern\'andez Santisteban}
%\affil{Anton Pannekoek Institute for Astronomy\\
% University of Amsterdam, Postbus 94249\\
%  1090 GE Amsterdam, The Netherlands}






\section*{} 

The All Sky Automated Survey for SuperNovae \citep[ASAS-SN;][]{shappee2014the} reported a possible Galactic dwarf nova ASASSN-18fs on 2018 March 19 at $\sim$13.2 mag in the V band, with a quiescent magnitude of V$>$17.6 \citep{stanek2018asassn}. Dwarf novae are accreting white dwarfs that occasionally show outbursts due to thermal instabilities in the accretion disk \citep[see][for a review]{lasota2001disc}. Here we report on the follow-up photometry using the {\it Neil Gehrels Swift Observatory}. 

ASASSN-18fs was observed using the Ultraviolet and Optical Telescope (UVOT) and the X-ray Telescope (XRT) aboard {\it Swift}. The data were downloaded from the HEASARC archive\footnote{https://heasarc.gsfc.nasa.gov/cgi-bin/W3Browse/swift.pl}. The source was observed 11 times since then, up to $\sim$46 days after the initial detection of the transient. 

\begin{figure}[htp!]
\centering
\includegraphics[width=0.95\textwidth]{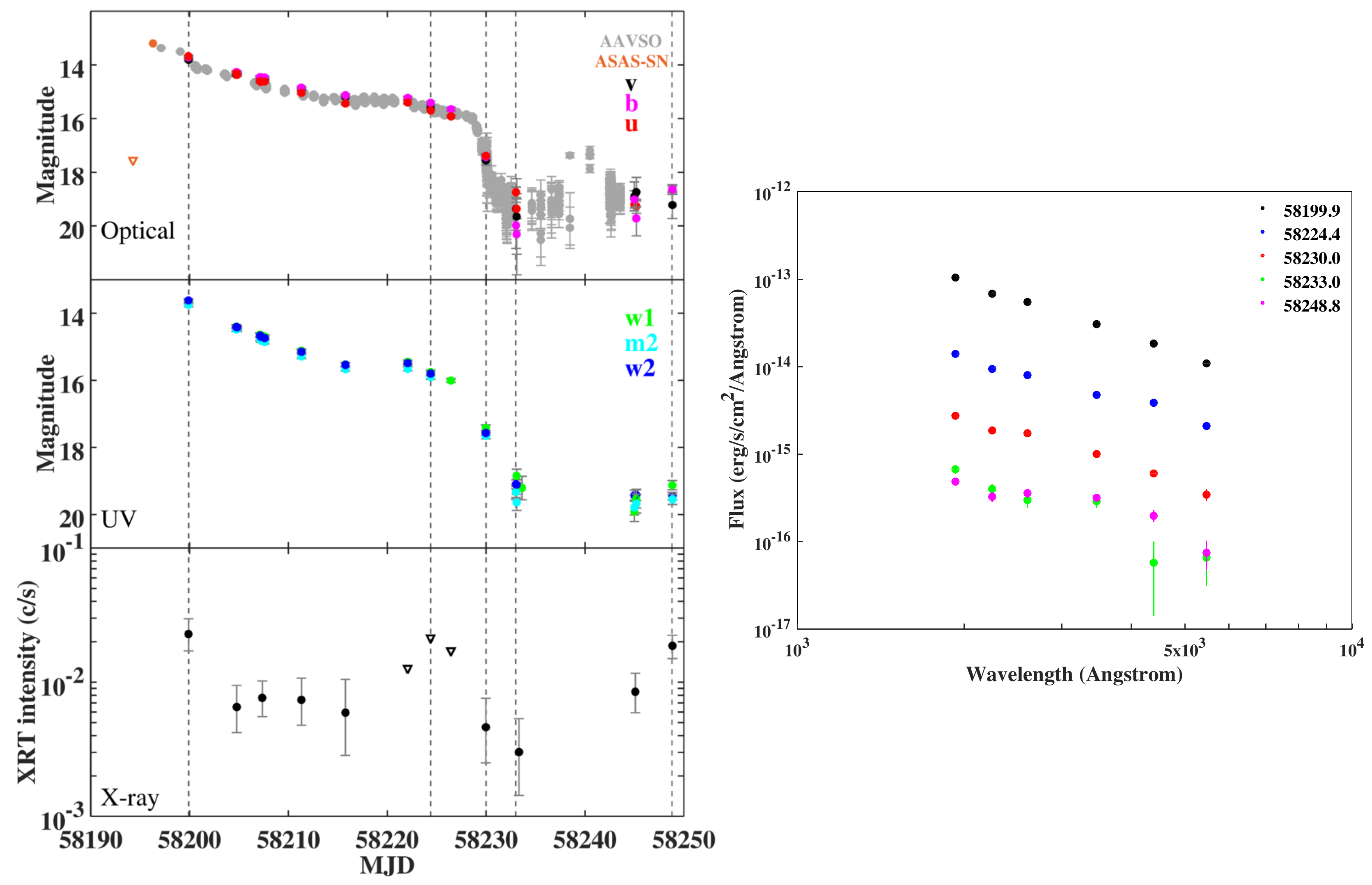}

\caption{\footnotesize
{The optical, UV, and X-ray light curves of ASASSN-18fs are shown in the upper, middle, and lower panels of the left figure, respectively. The vertical dashed lines indicate the times for which we constructed the SEDs (shown in the right figure).}}
\label{fig_imag}
\end{figure}

The UVOT products were analysed using \texttt{uvotsource} which extracted the source magnitude and flux. We used a circular source region of 5$^{\prime\prime}$ centred on the source location and a background region consisting of three 10$^{\prime\prime}$ circles placed near the source. The UVOT light curves are shown in Figure \ref{fig_imag} (left). The outburst lasted for $\sim$35 days, with the magnitude decaying from $\sim$13.7 to $\sim$15.8 mag during the initial $\sim$29 days. The source magnitude then decayed more rapidly over the next $\sim$6 days before reaching quiescence. %Further observations, $\sim$12 days and $\sim$15 days later showed that the source was still at the same magnitude. %This further confirmed that the source had reached quiescence. 

Several dwarf novae exhibit rebrightenings after their main outburst \citep[e.g.,][]{kato2009survey}. Due to the large gap in our {\it Swift} observations around MJD 58240, we were unable to constrain with these data whether or not the source exhibited rebrightenings. The ASAS-SN light curve\footnote{https://asas-sn.osu.edu/} around this time (not shown in the figure) was also non-constraining, providing only upper limits. Therefore, we examined archival data from the American Association of Variable Star Observers \citep[AAVSO\footnote{Using the CV band for which the unfiltered data are reduced to the V band}; Figure \ref{fig_imag}, left, grey points;][]{kafka2018aavso} which shows that ASASSN-18fs exhibited at least one post-outburst rebrightening. %Since the AAVSO coverage ended at MJD 58244, we could not determine if the source exhibited further rebrigthenings.

%
%\begin{figure}[htp!]
%\begin{center}
%\hbox{
%
%\includegraphics[width=0.5\textwidth]{ASASSN_18fs_lc}
%
%\hspace{-0.3cm}
%
%\includegraphics[width=0.55\textwidth]{asassn18fs_sed}
%}
%\end{center}
%\vspace{-0.5cm}
%\caption{\footnotesize
%{The optical, UV, and X-ray light curves of ASASSN-18fs are shown in the upper, middle, and lower panels of the left figure, respectively. The vertical dashed lines indicate the times for which we constructed the SEDs (shown in the right figure).}}
%%{{\it Left}: The optical, UV, and X-ray light curves of ASASSN-18fs are shown in the upper, middle, and lower panels, respectively. The vertical dashed grey lines indicate the times for which the SEDs were constructed.  {\it Right}: The SEDs of ASASSN-18fs using the UVOT data at various times during the outburst, during the transition to quiescence, and during quiescence, are shown.}}
%\label{fig_imag}
%\end{figure}

We also examined the Spectral Energy Distribution (SED; Figure \ref{fig_imag}, right) constructed using the {\it Swift}/UVOT data. The days for which these SEDs were constructed are shown by the vertical dashed lines in Figure \ref{fig_imag} (left). The SED evolution is consistent with that of a dwarf nova in outburst whose emission is dominated by a disk decaying with time.

The XRT data were processed using \texttt{xrtpipeline}. The light curve and spectra were extracted using \texttt{XSpec} (v12.9) using a circular source extraction region having a radius of 30$^{\prime\prime}$, centred on the source location. An annular background region having an inner and outer radius of 100$^{\prime\prime}$ and 200$^{\prime\prime}$, respectively, was used. The 0.5--10 keV light curve is shown in Figure \ref{fig_imag} (left bottom). The first XRT pointing detected the source at a count rate of 2.3$\times$10$^{-2}$ counts s$^{-1}$. By the next observation ($\sim$5 days later) the count rate had dropped by a factor of $\sim$3.5. It stayed at this count rate for the rest of the outburst and the subsequent decay. However, once the source transitioned to quiescence (after the rebrightening) the count rate was seen to rise, once again. It reached a similar level as was seen during the first XRT pointing. This behaviour, of the suppression of the X-ray emission during the outburst, is seen in several dwarf nova systems \citep[e.g., SS Cygni;][]{mcgowan2004on}.

\section*{}
\noindent We acknowledge the use of data from the {\it AAVSO International Database}.

 \newcommand{\noop}[1]{}

\end{document}